\documentstyle[11pt,nature]{article}
\begin{document}
\heading{Galaxy Formation at High Redshift} 

\author{N. Metcalfe, T. Shanks, A. Campos, R. Fong \& J.P. Gardner,}
{Physics Dept., Univ. of Durham, South Road, Durham, DH1 3LE, U.K.}

\noindent{\bf Sensitive optical surveys have revealed$^1$ a large population
of `faint blue galaxies' which are believed to be young galaxies$^2$,
observed close to their time of formation$^2$. But there  has
been  considerable uncertainty regarding the epochs at which these galaxies
are observed, owing to the difficulties inherent in determining 
spectroscopic redshifts for very faint objects. Here,
by modelling  the counts and colours of  galaxies at the faintest
detection limits,  we show that  the faint blue  galaxies are  likely to 
lie at high redshift  ($z\approx2$). This conclusion holds regardless
of whether the Universe is assumed to be open or at the critical
density (flat). In an open universe, the data are consistent with
galaxy models in which star formation rates decay exponentially with
decreasing redshift, whereas the assumption of a flat universe requires the
addition of a population of galaxies which are seen only at high redshift.}

We have  measured  galaxy counts and colours in the Deep Field$^{3}$
recently observed by the Hubble Space Telescope (HST)  to $U=27^m$,
$B=29^m$, $R=28.^m5$ and $I=28^m$. The HST data  covers  $\rm 5.3
arcmin^2$ and the exposure times were $\approx 25$ hrs in each of the 
BRI bands and $\approx47$hrs in the U band. Similar techniques, 
adjusted for the smaller image size, were used to analyse the HST
data  as we have previously used to analyse deep ground-based
images$^{4,5}$. Taking these deepest galaxy count data together  with
the results of recent galaxy redshift surveys at the spectroscopic
limit of the Keck Telescope$^{6,7}$, reveals  the clearest picture so
far of the Universe at faint magnitudes.

The derived  HST B galaxy number counts are shown in Fig. 1, together
with those from  other work, including the deepest ground-based counts
from the William Herschel Telescope (WHT) to $B=28.^m2$.  The HST and
WHT results are seen to be in good agreement to this  limit,  with the
HST counts extending $\approx1^m$ fainter.  Figure 1 also   shows the
HST counts in the I band. In this band  the HST data extends about
$10\times$ deeper than previous  data, because of the higher HST
resolution and the fainter background sky. Although HST has as yet no
K band imaging capability we also summarise in Fig. 1 the deepest
ground-based K counts, including new data from the U.K. Infrared
Telescope (UKIRT). The K counts  now appear to be reasonably  well defined in
the range $13^m<K<24^m$. 

We first compare the count data with non-evolving models. In the
modelling we assume throughout a Hubble Constant of 
$H_o=50kms^{-1}Mpc^{-1}$ but  our
main conclusions below are insensitive to this assumption. As can be seen
from Fig. 1, the non-evolving models increasingly  underestimate the
optical counts at faint magnitudes. However, the
magnitude at which the `faint blue galaxy' excess  becomes apparent is
dependent on the normalisation of the models  at bright  magnitudes. With
the high normalisation adopted here, non-evolving models  give a
reasonable representation of  the bright counts and redshift
distributions  in the range $18^m<B<22.^m5^{1,4,5,8,9}$ and also to
individual counts of spiral and early-type galaxies at similar
magnitudes$^{10,11}$. As a consequence, the B galaxy count then only
shows evidence for strong evolution  in the range $B>23^m$. Also the high
normalisation  allows  non-evolving models with deceleration parameter
$0.05<q_o<0.5$, to fit  the K counts from $K=15^m$ to the faintest count  
limit at $K=24^m$. This is as expected since evolved, young star
populations are expected to affect galaxy light less in the infrared than
in the blue.

To fit the optical counts, we then consider simple evolutionary
models from Bruzual \& Charlot$^{12}$ where galaxy  star-formation rates
rise exponentially with look-back time.   In such models, the galaxy
light at early times is generally dominated by luminous, blue stars but
at later times, when these  stars fade and the star-formation rate slows,
the  galaxy light dims and reddens. These models  are known to fit the B
counts in the range $18^m<B<25^m$ $^{1,4,5,8,9}$.   But previous faint
galaxy redshift surveys at $B<24^m$ presented a problem for such models,
as they  predict a high  redshift tail of evolved, luminous galaxies which
was unobserved in these surveys$^{13}$. However, these  surveys were
frequently  only $\approx$60\% complete and the galaxies with
unidentified  redshifts were usually blue. Recently Cowie et al$^{6,7}$
have used the Keck 10m telescope to make a new $B<24^m$ galaxy redshift
survey with $>80$\% completeness and have detected such a high redshift
galaxy component, supporting the basic viability of these models (see
Fig. 2). An extended high redshift tail is also consistent with the low
galaxy clustering amplitude observed$^9$ at $B>23^m$.

A further new  development is that the latest version of the preferred 
spiral galaxy evolution  model$^{12}$  now permits more  brightening  at
high redshift,  while still reproducing the   blue   colours  of spirals
at the present day$^{14}$.  Combined with the  steep luminosity function
of spirals$^{8}$ and assuming a small amount of internal
dust$^{14-16}$, this enables us to obtain a good fit  to the  high
redshift tail of the Keck, number-redshift distribution, n(z), at
$22.^m5<B<24^m$ (see Fig. 2) and to other redshift survey results$^{17}$.
In the low $q_o$ case, this spiral dominated  model then also produces a
reasonable fit to the optical counts to $B\approx27^m$ and $I\approx26^m$
(see Fig. 1).   Although the model underestimates the optical counts at
fainter magnitudes,  this discrepancy is probably still within the combined
data and  model uncertainties. In the $q_o=0.5$ case,  the spiral luminosity
evolution model only fits the optical data  to $B\approx25^m$ and
$I\approx23.^m5$ and  then more seriously underestimates the counts at
fainter magnitudes.  Thus, the HST data confirms the previous ground-based
result$^{1,4,5,18}$ that if $q_o=0.5$, then there is not enough spatial
volume at high redshifts to allow simple luminosity evolution models to
fit the high galaxy counts at $B>25^m$. This confirmation is important
because confusion corrections  are much smaller in the high resolution
HST galaxy counts than in previous ground-based data.

We have already noted that the faint K counts seem less affected by
evolution and this  is supported by the new Keck redshift 
data$^{6,7}$, where non-evolving models again give a good fit to  the
n(z) relation  for $18^m<K<19^m$ (ref. 7). Since  non-evolving models imply
that, at  $K<20^m$,  the majority of  galaxies are early-types, this
suggests that  early-type galaxies may be little affected by  either
dynamical(merging) {\it or} luminosity evolution.  Indeed, although the
effect of passive evolution of early-type galaxies with a Salpeter Initial
Mass Function (IMF) slope (x=1.35) is small at K($<0.7^m$ at $z=1$), it is
still big enough to make the predicted Keck n(z) appear too extended at
$18^m<K<19^m$$^{6,7}$.  If the result is not due to incompleteness, then
the amount of K evolution can be reduced to acceptable limits by assuming
a  more dwarf dominated IMF(x=3) for early-type galaxies. While awaiting
more complete K surveys, we have adopted  this dwarf-dominated IMF for
early types throughout this paper.

To improve the fit of the  $q_o=0.5$ model to the faint optical counts 
we consider a  model with an extra population of high redshift galaxies
which have a constant star-formation rate from the formation epoch till
z=1 ($\approx4Gyr$ after formation with our assumed $H_o$). The
Bruzual \& Charlot$^{12}$ model shows that at  $z\la 1$ the galaxy
then rapidly fades by 5 magnitudes in B to  form a red  (dE) galaxy by
the present day. This model is in the spirit of previously proposed
`disappearing dwarf' count models$^{19}$ and it gives  a good fit to
the faint B,I,K counts  and the Keck n(z) data (see Figs. 1 and 2). 

We next test these models against the faint galaxy colour distributions
in the Hubble Deep Field. The presence of broad  features  in galaxy
spectra allows tests to be made of the predicted galaxy redshifts. One
approach is to use the HST broad band photometry as a rough galaxy
spectrum and then derive redshifts for individual galaxies using
local galaxies as templates$^{20}$. Here we follow the different
approach of  simply comparing  our evolutionary model predictions to the 
observations of faint galaxy colours.  We see in Fig.3 that our  predicted
$U'-B':B'-I'$ model tracks, ie the loci traced by model galaxy colours as
they change with redshift, compare well with the observed colours for
$B'<27.^m5$ galaxies. (Primed letters here denote the natural HST magnitude
system). In particular, the redshifting of the Lyman $\alpha$
forest/break$^{21}$ absorption features through the $U'$ band causes the
model $U'-B'$ colours to move sharply redwards at $B'-I'\approx0.^m7$ and
the same effect is clearly seen in the data. The colours of 45 brighter
$B\approx24^m$ galaxies with Keck redshifts  are also shown in Fig. 3 and
are also found to  agree well  with their predicted colours. (The
equivalent $B'-R':R'-I'$ graph is shown as supplemental information at the
Nature web site). 

The above predictions show that, for the majority of faint
galaxies,  $U'-B'<0$ is predicted to correspond to $z<2$ galaxies and
$U'-B'>0$ corresponds to $z>2$ galaxies. We find that the proportion of
galaxies with $U'-B'>0$ (including those undetected in $U'$) rises to
$47\pm7$\% of the total at $27^m<B<28^m$, indicating that the redshift
distribution may peak at $z\approx2$. This fraction is matched
very well by both the $q_o=0.05$ model  which predicts 47\% with
$U'-B'>0$ at the same limit and the $q_o=0.5$, disappearing dwarf (dE) model
which  predicts 43\%. We have also considered another $q_o=0.5$ model 
which assumes  an extra population of low redshift dwarf spirals (dSp)
which evolve more slowly according to our standard exponential model
for spiral luminosity evolution$^{12}$. Although this model also gives an
improved fit to the counts, it predicts too few  high redshift
($U'-B'>0$) galaxies (28\%) for compatibility with the faint HST data.

These conclusions are confirmed by consideration of the  $B'-R':R'-I'$
colour-colour plot in Fig. 4. Fig. 4a shows  the predicted tracks of
the galaxy types with redshift, as in Fig. 3.  Also  plotted are the
galaxies with $U'-B'>0$ and $R'<27.^m5$; these are expected to have 
$z>2$ by the above arguments and it can be seen that their position on
the $B'-R':R'-I'$ tracks is entirely consistent with their lying in this
redshift range. We regard this as  crucial confirmation that our
models are indicating consistent redshifts for galaxies in $U'-B'$  and 
$B'-R':R'-I'$  independently. Figs. 4b,c,d then show the HST data (dots)
at $R'<28^m$ compared to the predicted galaxy number contours for the
open and closed models, based on the tracks shown in Fig. 4a. Both the
$q_o=0.05$ and the $q_o=0.5$ dE model contours give a reasonable fit
to the data which seem to peak at $B'-R'\approx0.3$, $R'-I'\approx0.3$,
corresponding to $z\approx2$ for all galaxy types. However, the
$q_o=0.5$ dSp model contours peak away from this point at
$B'-R'\approx1$, $R'-I'\approx1$ which corresponds to $z\approx0.5$ for 
the dwarf spiral galaxies and we  conclude that the galaxy redshift
distribution in this model is skewed to too low redshifts to be 
compatible with the colour data. This does not mean that the above
`disappearing dwarf' model is unique in allowing a fit to be obtained
with  $q_o=0.5$; other possibilities such as  merging models may also
exist. However, it does suggest that, in any model, the star-forming
phase has to be at $z\approx2$ for consistency with the faint
galaxy colours in the Hubble Deep Field.

\newpage
   
\references

\natref{1}{Koo, D.C. \& Kron, R.G.} {ARA\&A}{{\bf 30}, 613-652}{(1992 and
references therein)}

\natref{2} {Cowie, L.L.}{In {\em `The Post-Recombination 
Universe'}, (eds Kaiser, N. \& Lasenby, A.),} {Kluwer,
Dordrecht, 1--18}{(1990)}

\natref{3}{Williams, R.E.{\it et al}} {{\em Astron. J.}} {in press}
{(1996)}

\natref{4}{Metcalfe, N., Shanks, T., Fong, R. \& Jones, L.R.}{\mn}
{{\bf 249}, 481-497}{(1991)} 

\natref{5}{Metcalfe, N., Shanks, T., Fong, R., \& Roche, N.} 
{\mn}{{\bf 273}, 257-276}{(1995 and references therein)}

\natref{6}{Cowie, L.L., Hu, E.M., \& Songaila, A.}
{\nat }{{\bf 377}, 603-605}{(1995)}

\natref{7}{Cowie, L.L., Songaila, A., Hu, E.M. \& Cohen, J.G.}
{{\em Astron. J.} }{in the press}{(1996)}

\natref{8}{Shanks, T.}{in {\em `The Galactic and Extragalactic
Background Radiations'}, (eds Bowyer, S. \& Leinert, C.),}{Kluwer,
Dordrecht, 269--281}{(1990)}

\natref{9}{Roche, N., Shanks, T., Metcalfe, N. \& Fong,  R.} 
{\mn}{{\bf 263}, 360-368}{(1993)}

\natref{10}{Glazebrook, K., Ellis, R.S., Santiago, B. \& Griffiths,
R.}  {\mn}{{\bf 275}, L19-L22}{(1995)}

\natref{11}{Driver, S.P., Windhorst, R.A., Ostrander, E.J., Keel, W.C.,
Griffiths, R.E.  \& Ratnatunga, K.U.} {\aj }{{\bf 449 },
L23-L28}{(1995)}

\natref{12}{Bruzual,  A.G. \& Charlot, S.}  {\aj} {{\bf 405},538-553}
{(1993)}

\natref{13}{Glazebrook, K., Ellis, R., Colless, M. Broadhurst, T., 
Allington-Smith, J.R. \& Tanvir, N.R.}{\mn}{{\bf 273},157-168}
{(1995)}

\natref{14}{Campos, A. \& Shanks, T.}{\mn}{{submitted},}{(1996)}

\natref{15}{Wang, B.}{\aj}{{\bf 383}, L37-L40}{(1991)}

\natref{16}{Gronwall, C. \& Koo, D. C.} {\aj}{{\bf 440}, L1-L4}{(1995)}

\natref{17}{Lilly, S.J., Tresse, L., Hammer, F., Crampton, D. \& Le
Fevre, O.}  {\aj} {{\bf 455},108-124}{(1995)}

\natref{18}{Yoshii, Y. \& Takahara, F.}  {\aj} {{\bf326},1-18}{(1988)}

\natref{19}{Babul, A. \& Rees, M.J.}  {\mn} {{\bf 255},346-350}{(1992)}

\natref{20}{Lanzetta, K.M., Yahil, A. \& Fernandex-Soto, A.}  {\em
Nature} {{\bf 381},759-763}{(1996)}

\natref{21}{Madau, P.}  {\aj} {{\bf 441},18-27}{(1995)}

\natref{22}{Smail, I., Hogg, D.W., Yan, L. \& Cohen, J.G.}
{\mn}{in press }{(1996)}

\natref{23}{Driver, S.P., Phillipps, S., Davies, J.I., Morgan, I. \&
Disney,  M.J.}{\mn}{{\bf 268}, 393-404}{(1994)}

\natref{24}{Djorgovski, S. et al}{\aj}{{\bf 438},L13-L16}{(1995)}

\natref{25}{Gardner, J.P., Cowie, L.L. \& Wainscoat, R.J.} 
{\aj}{{\bf 415}, L9-L12}{(1993)}

\natref{26}{Soifer, B.T. {\it et al}} {\aj }{{\bf 420},L1-L4}{(1994)}

\natref{27}{McLeod,  B.A., Bernstein, G.M., Rieke, M.J., Tollestrup,
E.V. \& Fazio, G.G.}{\em ApJ Supp. }{{\bf 96},117-121}{(1995)}

\natref{28}{Glazebrook, K., Peacock, J.A., Collins, C.A. \& Miller,
L.,} {\mn}{{\bf 266}, 65-91}{(1995)}

\natref{29}{Gardner, J.P., Sharples, R.M., Carrasco, B.E. \& Frenk, C.S.,
}  {\mn}{submitted}{(1996)}

\newpage

\acknowledgements
We gratefully acknowledge  L.L. Cowie for allowing us to use results from the
Keck 10m redshift surveys in advance of publication. We also gratefully
acknowledge A.G. Bruzual for producing dwarf dominated evolutionary models
specially for this paper. We also acknowldge the use of  the Hubble Space
Telescope Deep Field data. A. Campos acknowledges receipt of an EC Fellowship.
N. Metcalfe and J.P. Gardner acknowledge PPARC funding. We thank
the referees for helpful comments.

A. Campos is now at Instituto de Astrofisica de Andalucia, CSIC, Spain.

\newpage

\begin{flushleft}
{\bf Figure 1.}
\end{flushleft}
The B (Johnson) and I (Kron-Cousins) galaxy counts from  the HST Deep
Field (F450W, F814W)   data$^3$ compared to  counts from the WHT, UKIRT and
elsewhere ( refs. 1,4,5,10,11,22-29) and  various models. Also shown are
deep K band galaxy counts from ground-based data. The I band counts have
been multiplied  by a factor of 10 and the K band counts have been
mutiplied by a factor of 100 for clarity. (The  HST U and R counts are
presented as supplemental information at the Nature web site.) The model
luminosity function parameters and other details of the modelling procedure
are given elsewhere$^{4,5}$.  The galaxy luminosity evolution with redshift
is computed from Bruzual \& Charlot$^{12}$ isochrone synthesis models, 
using  the appropriate passbands.  We assume  galaxy ages of 16 and 12.7
Gyr in the cases $q_o=0.05, 0.5$. We adopt a dwarf dominated IMF (x=3) 
with an exponentially increasing star-formation rate of time scale,
$\tau=2.5Gyr$, for E/SO/Sab  galaxies and a Salpeter IMF (x=1.35) with
an exponentially increasing star-formation rate of time scale,
$\tau=9Gyr$, for Sbc/Scd/Sdm  galaxies. The latter  are also assumed to
have  internal dust absorption at $z=0$ of $A_B=0.^m3$, $A_I=0.^m11$,
$A_K=0.^m03$. We take an absorption laaw inversely proportional to the
wavelength $\lambda$, $A_\lambda\propto 1/\lambda$ to approximate the effect
of redshift on the internal dust absorption. The models also include the
effect of  Lyman $\alpha$ forest/break absorption$^{23}$. Spiral evolution
dominates these models in the  B and I bands. The $q_o=0.05$ evolving model
gives a good fit to $B\approx27^m, I\approx26^m$,  whereas the $q_o=0.5$
model only fits to $B\approx25^m, I\approx23.^m5$. The fit of the $q_o=0.5$
model is  improved  when an extra high redshift galaxy population (dE) with
constant star-formation rate  at $z>1$  and rapidly fading at $z<1$ is
invoked. The Schechter luminosity function parameters of the dE population
at z=0 are ${M^*}_B=-16^m.0$, $\alpha=-1.2$ and
$\phi^*=0.019$mag$^{-1}$Mpc$^{-3}$. The K galaxy counts, in contrast to the
B and I counts, are well fitted by non-evolving models.

\begin{flushleft}
{\bf Figure 2.}
\end{flushleft}

Thegalaxy number-redshift distribution, n(z), for $22.^m5<B<24^m$
implied by new redshift data acquired on the Keck Telescope$^{6,7}$.
The observed n(z) is clearly more extended than the non-evolving models
with either $q_o=0.05$ or $q_o=0.5$. The extended redshift  distribution is
well fitted by our evolutionary  models  whose parameters are described in
the Fig. 1 legend.

\begin{flushleft}
{\bf Figure 3.}
\end{flushleft}

Dots represent the $U'-B':B'-I'$ colours of $B'<27.^m5$ galaxies in the
Hubble Deep Field. Primed letters for magnitudes indicate that here we
are using the natural HST magnitude system, with zeropoint set at an A0V
star. The arrows represent detection upper limits, mainly  galaxies which
are undetected in $U'$. The $U'-B'$ colours move sharply redwards at
$B'-I'\approx0.8$ due to the Lyman $\alpha$ forest/Lyman break passing
through the $U'$ band.   The predicted tracks are the $q_o=0.05$
evolutionary models for each morphological type as detailed in the
caption to Fig. 1,  modulated in the case of Sbc/Scd/Sdm types by our
assumed internal dust absorption of $A_U'=0.^m45$, $A_B'=0.^m3,
A_I'=0.^m11$ and  in the case of all galaxies by the Lyman $\alpha$
forest absorption.  The models used in the $q_o=0.5$ case (not shown)
show a very  similar behaviour, even for the rapidly fading dE type. The
z=1 and z=2 labelled positions on the tracks indicate the colours of model
E/SO and Sdm galaxies at these redshifts. The remaining symbols represent
the colours of 45 brighter galaxies with Keck spectroscopic redshifts are
also shown and these agree well with the predicted colours for these
galaxies. It can also be seen that $U'-B'<0$ is predicted to correspond to
galaxies with $z<2$ and $U'-B'>0$ to galaxies with $z>2$.

\begin{flushleft}
{\bf Figure 4a.}
\end{flushleft}

The $q_o=0.05$ evolutionary models' $B'-R':R'-I'$ tracks with redshift, 
as a function of galaxy morphological type.  Primed letters for
magnitudes  indicate that here we are using the natural HST magnitude
system, zeropointed to an A0V star. The tracks are modulated by our
assumed internal dust and Lyman $\alpha$ forest/break absorption. A
filled symbol marks each galaxy type's colour at zero redshift. The open
symbols mark the galaxy type's colour at unit intervals in redshift for
E/SO and Sdm galaxies. Galaxies with $B'-R'=0.^m3$ and $R'-I'=0.^m6$ have
$z\approx2$. The same is true for the $q_o=0.5$ models,  even for the
rapidly fading dE type (not shown). The dots indicate the colours of
galaxies with $R'<27.^m5$ and $U'-B'>0$ which are predicted  to have
$z>2$.  Their position in the independent $B'-R':R'-I'$ plane is consistent
with this prediction.

\begin{flushleft}
{\bf Figure 4b.}
\end{flushleft}

Dots represent the $B'-R':R'-I'$ colours of $R'<28^m$ galaxies in the
Hubble Deep Field. These show the same horse-shoe shaped track as
expected from the models in Fig. 4a. The contours represent the relative
numbers of galaxies predicted for the $q_o=0.05$ evolutionary model,
which appear to be in good agreement with the data, peaking at the
colours corresponding to $z\approx2$.

\begin{flushleft}
{\bf Figure 4c.}
\end{flushleft}

As Fig. 4b with the contours here representing the relative numbers
of galaxies predicted for the $q_o=0.5$ `disappearing dwarf' (dE) model,
which again  appear to be in reasonable agreement with the data.

\begin{flushleft}
{\bf Figure 4d.}
\end{flushleft}

As Fig. 4b with the contours here representing the relative numbers
of galaxies predicted for the $q_o=0.5$ low redshift dwarf (dSp) model. The
data show that this model is inappropriate, as it clearly predicts too
many low redshift, $z\approx0.5$, spiral galaxies.

\end{document}